\documentclass[superscriptaddress,amsmath,amssymb,
aps, prstab, twocolumn, groupedaddress]{revtex4-2} 
\usepackage{amsmath}
\usepackage{natbib}
\usepackage{float}
\usepackage{amssymb}
\usepackage{physics}
\usepackage{dcolumn}
\usepackage{graphicx}
\usepackage{longtable}
\usepackage{bm}
\usepackage{xcolor}
\usepackage{epsfig}
\usepackage{xr-hyper}
\usepackage[colorlinks=true, citecolor=blue,linkcolor=blue, urlcolor=blue]{hyperref}
\usepackage{enumerate}

\begin{document}
\raggedbottom

\title{
Low temperature thermodynamics of $S_{\mathrm{eff}}=1/2$ triangular lattice  quantum spin liquid candidate TlYbS$_2$}

\author{Bishnu P. Belbase$^{1}$}
\thanks{\text{These authors contributed equally to this work}}
\author{Arjun Unnikrishnan$^{1,2,3, \ast}$}
\email{arjunpookad@gmail.com}
\author{Piyush Chhallare$^{1}$}
\author{Mohan B. Neupane$^{1}$}
\author{Eun Sang Choi$^{4}$}
\author{Sebastian Erdmann$^{2}$}
\author{Muhammad U. Akbar$^{5}$}
\author{Mamoun Hemmida$^{5}$}
\author{H.-A. Krug von Nidda$^{5}$}
\author{Philipp Gegenwart$^{2}$}
\author{Arnab Banerjee$^{1}$}
\email{arnabb@purdue.edu}

\affiliation{$^1$Department of Physics and Astronomy, Purdue University, West Lafayette, Indiana 47907, USA}
\affiliation{$^2$Experimental Physics VI, Center for Electronic Correlations and Magnetism, Institute of Physics, University of Augsburg, Augsburg 86159, Germany}
\affiliation{$^3$Solid State and Structural Chemistry Unit (SSCU), Indian Institute of Science, Bengaluru - 560012, India}
\affiliation{$^4$National High Magnetic Field Laboratory, Tallahassee, Florida 32310, USA}
\affiliation{$^5$Experimental Physics V, Center for Electronic Correlations and Magnetism, Institute of Physics, University of Augsburg, Augsburg 86159, Germany}
\date{\today}

\begin{abstract}
Geometrically frustrated triangular-lattice antiferromagnets exhibit a delicate competition between magnetic order and quantum spin liquid (QSL) behavior, with the Yb-based delafossite family $A$Yb$X_2$ providing a structurally clean platform for exploring this physics. Here, we report a comprehensive study of single-crystal TlYbS$_2$ using DC magnetization, AC susceptibility, electron spin resonance (ESR), and specific heat measurements extending from room temperature to the millikelvin regime. Single-crystal X-ray diffraction confirms a trigonal $R\bar{3}m$ structure comprising well-separated triangular layers of Yb$^{3+}$ ions with no detectable site disorder. At zero field, a weak thermodynamic anomaly is observed near $530$~mK, which may indicate the onset of a weakly ordered state similar to that reported in KYbSe$_2$. Below $300$~mK, the zero-field magnetic specific heat follows $C_{\rm m}\propto T^{1.8}$, close to a quadratic temperature dependence, in contrast to the linear temperature dependence reported for the sister compound TlYbSe$_2$, which has been described in terms of the interplay between spinons and thermally excited gauge-flux excitations. Magnetization and ESR measurements establish pronounced easy-plane magnetic anisotropy, with $g_\perp/g_\parallel \approx 6.9$. The anomaly near $530$~mK exhibits a strongly anisotropic response to magnetic field. For $H\perp c$, it remains visible up to approximately $2$~T and continuously evolves toward a field-induced ordered phase above approximately $2.5$~T, followed by a sequence of field-induced phases that includes a $1/3$ magnetization plateau between approximately $5$ and $8$~T and full polarization near $17$~T. In contrast, for $H\parallel c$, the anomaly weakens and is suppressed near $3$~T, consistent with the recently reported confinement--deconfinement transition from an ordered state to a field-induced QSL. 
\end{abstract}

\maketitle

\section{Introduction}
Geometrically frustrated magnets provide an ideal platform in which strong quantum fluctuations can suppress conventional magnetic order and stabilize unconventional ground states. Among these, the spin-$1/2$ triangular-lattice antiferromagnets (TLAFs) serve as paradigmatic model systems. In the ideal nearest-neighbor (NN) Heisenberg limit, the ground state is expected to exhibit a $120^\circ$ magnetic order; however, additional interactions, such as next-nearest-neighbor (NNN) exchange and exchange anisotropy, can enhance quantum fluctuations, destabilize this order, and promote quantum-disordered phases, including quantum spin liquids (QSLs)~\cite{bernu1994exact,capriotti1999long,iqbal2016spin,zhu2015spin}. These states are characterized by the absence of long-range magnetic order down to the lowest temperatures, persistent spin dynamics, and the emergence of fractionalized excitations.

Rare-earth-based triangular-lattice compounds have emerged as particularly promising realizations of this physics. In these materials, strong spin--orbit coupling and crystal electric field effects isolate a Kramers doublet ground state, allowing the low-energy magnetism to be described by an effective spin-$1/2$ Hamiltonian with pronounced exchange anisotropy. The delafossite family $A$Yb$X_2$ ($A$ = monovalent ion, $X$ = O, S, Se) provides an ideal platform in this regard, as Yb$^{3+}$ ions form well-defined triangular layers with minimal magnetic-site disorder~\cite{scheie2024proximate,ranjith2019anisotropic,dai2021spinon,bordelon2019field,ding2019gapless,baenitz2018naybs,belbase2026finite,khatua2024magnetic,guo2019magnetism,ferreira2020frustrated,khattar2025magnetic}, providing relatively clean platforms compared to some other Yb-based candidates~\cite{paddison2017continuous,zhu2017disorder}.

One of the most intriguing aspects of the delafossite $A$Yb$X_2$ family is the remarkable tunability of its magnetic ground state. Chemical substitution at the $A-$ or $X-$site modifies the competing magnetic interactions, including the ratio of next-nearest-neighbor to nearest-neighbor exchange, $J_2/J_1$, thereby tuning the system across the boundary between conventional $120^\circ$ order and quantum-disordered regimes~\cite{scheie2024nonlinear}. In this context, NaYbO$_2$, NaYbS$_2$, and NaYbSe$_2$ exhibit persistent spin dynamics and evade conventional long-range order down to very low temperatures~\cite{bordelon2019field,baenitz2018naybs,ranjith2019anisotropic}. KYbSe$_2$ develops weak $120^\circ$ magnetic order below $290$~mK, yet remains close to the proposed ordered-to-QSL boundary, with signatures of fractionalized spin excitations observed in neutron-scattering experiments~\cite{scheie2024proximate}. Replacing the alkali-metal with thallium introduces an additional route for tuning the magnetic interactions. In particular, TlYbSe$_2$ exhibits no long-range magnetic order down to $20$~mK despite relatively strong antiferromagnetic exchange interactions and displays a linear-in-temperature magnetic specific heat consistent with a finite low-energy spinon density of states~\cite{belbase2026finite}. Its proposed location between NaYbSe$_2$ and KYbSe$_2$ in the $J_2/J_1$ phase diagram, as shown in Fig.~\ref{HC_comparison}(b), suggests close proximity to the boundary separating the $120^\circ$ ordered phase from the QSL regime. The sulfur analog TlYbS$_2$ therefore offers an ideal opportunity to investigate how chalcogen substitution tunes the competing magnetic interactions and reshapes the low-energy excitation spectrum.

Recently, thermal-transport and nuclear magnetic resonance (NMR) measurements on TlYbS$_2$ for magnetic fields applied along the crystallographic $c$ axis ($H\parallel c$), using crystals from the same series as those studied in this work, revealed a field-induced confinement--deconfinement transition from an ordered state to a gapless QSL above $3$~T~\cite{hosoi2026continuous}. Here, we report a comprehensive investigation of single-crystalline TlYbS$_2$ using DC magnetization, AC susceptibility, electron spin resonance (ESR), and specific-heat measurements extending from room temperature to the millikelvin regime, with particular emphasis on fields applied perpendicular to the crystallographic $c$ axis ($H\perp c$). An analysis of the results suggests that TlYbS$_2$ occupies an intermediate regime between the weakly ordered and quantum-disordered members of the Yb-based delafossite family, providing a new reference point for understanding how competing exchange interactions and magnetic anisotropy shape the ground state across this family.

\section{Experiments} 

\begin{figure*}[!htb]
\centering
\includegraphics[width=\linewidth]{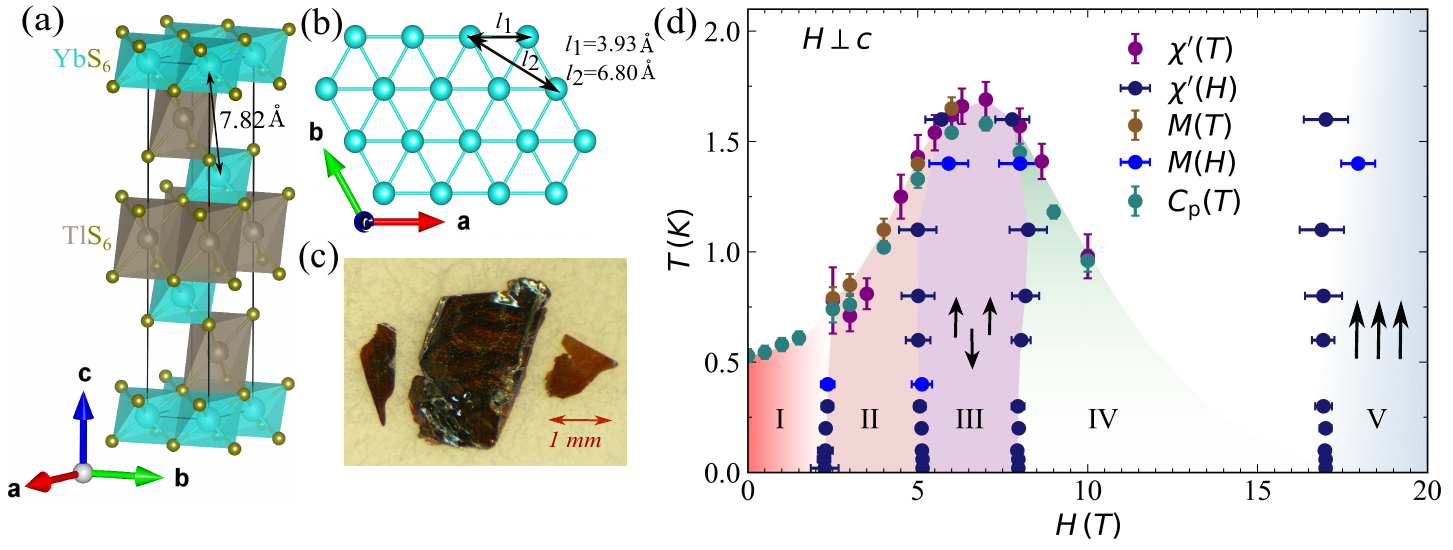}
\caption{(a) Crystal structure of TlYbS$_2$, highlighting the stacking of Yb-based triangular layers separated by Tl--S slabs along the crystallographic $c$ axis. 
(b) Top view of the $ab$ plane showing the two-dimensional triangular network of Yb ions. The nearest- and next-nearest-neighbor in-plane Yb--Yb distances are indicated. (c) Single crystals of TlYbS$_2$, showing their plate-like morphology.
(d) Temperature-field ($H-T$) phase diagram of TlYbS$_2$ constructed from AC susceptibility ($\chi'$), DC magnetization ($M$), and specific heat ($C_{\rm p}(T)$) measurements. Distinct magnetic regimes are highlighted by different shaded regions.}  
    \label{structure}
\end{figure*}

\subsection{Single Crystal Growth} 

Single crystals of TlYbS$_2$ were grown using a two-step procedure consisting of polycrystalline synthesis followed by flux growth. Polycrystalline TlYbS$_2$ was first prepared by a conventional solid-state reaction using stoichiometric amounts of thallium granules (99.99\%, Thermo Fisher), ytterbium powder (99.9\%, Thermo Fisher), and sulfur powder (99.999\%, Thermo Fisher) under vacuum, following the method reported in Ref.~\cite{ferreira2020frustrated}.
Single crystals were subsequently grown from this precursor using TlCl flux at a maximum temperature of $690~^\circ$C. The growth yielded ruby-red, plate-like single crystals with typical lateral dimensions of $\sim2~\times~2$~mm$^2$.

\subsection{X-ray Diffraction and Energy-Dispersive X-ray Spectroscopy} 

The phase purity of the polycrystalline material obtained from the solid-state reaction was verified by powder X-ray diffraction (XRD) at room temperature using a PANalytical Empyrean diffractometer equipped with an incident-beam monochromator and Cu~K$_{\alpha}$ radiation. 
Single-crystal X-ray diffraction measurements were performed at room temperature using a Bruker D8 Quest diffractometer with a Mo K$\alpha$ radiation source ($\lambda = 0.71073$ \AA). Diffraction data were collected using APEX6 with $\omega$- and $\phi$-scan techniques. The crystal structure was solved and refined using the \textsc{ShelX} software package~\cite{sheldrick2008short}. We performed energy-dispersive X-ray (EDX) spectroscopy using a Thermo Fisher Scientific Apreo $2$S FEG SEM equipped with an Oxford ULTIM Max $100$ detector at the Purdue Electron Microscopy Center (RRID:SCR$\_022687$) to determine the elemental composition of the single crystals. EDX data processing was performed using AZtec software (Oxford Instruments). To prevent charging during electron-beam exposure, the sample surface was coated with a thin palladium (Pd) layer prior to scanning electron microscopy (SEM) and EDX characterization.    

\subsection{ Magnetization Measurements}
DC magnetization measurements down to $1.8$~K in magnetic fields up to $7$~T were carried out using a Quantum Design SQUID magnetometer (MPMS$-$3) at the Birck Nanotechnology Center, Purdue University. These measurements were subsequently extended to lower temperatures down to $0.4$~K under magnetic fields up to $7$~T using a $^3$He insert in an MPMS$-$3 system at the University of Augsburg. High-field isothermal magnetization measurements up to $35$~T were performed using a vibrating sample magnetometer (VSM) at the National High Magnetic Field Laboratory (NHMFL).

AC susceptibility measurements were conducted at the NHMFL using two different superconducting magnet systems. Measurements down to $300$~mK were performed in the $18$~T General Purpose Superconducting Magnet (SCM2) equipped with a $^3$He insert, employing an AC susceptometer with an excitation frequency of $200$~Hz and an AC field amplitude of $2$~Oe. To access lower temperatures, additional measurements down to $20$~mK were carried out in an $18$~T superconducting magnet equipped with a top-loading dilution refrigerator (SCM1), using an AC field of $0.5$~Oe at a frequency of $200$~Hz.

\subsection{ESR}
The electron spin resonance (ESR) measurements were performed using a continuous-wave spectrometer (Bruker ELEXSYS E500) at X-band frequency ($\nu = 9.4$~GHz) in the temperature range $4 \leq T \leq 300$~K, employing a continuous He gas-flow cryostat (Oxford Instruments). ESR detects the power $P$ absorbed by the sample from the transverse magnetic microwave field $h_{\mathrm{mw}}$ as a function of the static magnetic field $H$, where $H \perp h_{\mathrm{mw}}$. 
The signal arises from magnetic dipole transitions between the Zeeman levels of the electron spins. The signal-to-noise ratio of the spectra is improved by recording the derivative $dP/dH$ using a lock-in technique with field modulation. The measurements were carried out on a single crystal fixed in a quartz tube with paraffin.

\subsection{Specific Heat}
Specific-heat measurements on TlYbS$_2$ and its nonmagnetic analog TlLuS$_2$ were first performed down to $1.8$~K using a Physical Property Measurement System (PPMS, Quantum Design) equipped with a $^4$He insert at the Spin-Lab facility in the Birck Nanotechnology Center, Purdue University. To extend the TlYbS$_2$ measurements to lower temperatures, additional measurements were performed at the University of Augsburg using a PPMS equipped with a $^3$He insert, covering temperatures down to $0.4$~K under various applied magnetic fields. Measurements were further extended down to $70$~mK using the relaxation method in a home-built setup installed in a dilution refrigerator at the University of Augsburg.

\section{Results} 

\subsection{Crystal Structure}

Single-crystal X-ray diffraction refinement confirms a trigonal crystal structure with space group $R\bar{3}m$ (No.~166), with no evidence of site mixing or crystallographic disorder within the sensitivity of the refinement. In addition, EDX measurements performed on multiple single crystals consistently yield a Tl:Yb:S ratio of $1:1:2$, confirming the expected stoichiometry. Crystallographic parameters obtained from the single-crystal refinement and elemental compositions from EDX measurements are summarized in the Supplementary Material (SM)~\cite{supp}. While earlier studies on polycrystalline samples assigned TlYbS$_2$ to the trigonal $R\bar{3}m$ ($\alpha$-NaFeO$_2$-type) structure~\cite{duczmal1994magnetic}, a previous single-crystal study by Ferreira \textit{et al.}~\cite{ferreira2020frustrated} proposed a hexagonal $P6_3/mmc$ ($\beta$-RbScO$_2$-type) structure. However, the experimental powder X-ray diffraction pattern reported by Ferreira \textit{et al.} is in much better agreement with the simulated pattern for the $R\bar{3}m$ structure than for the $P6_3/mmc$ structure. A detailed comparison is provided in the SM~\cite{supp}.

The structure consists of two-dimensional triangular layers of edge-sharing YbS$_6$ octahedra in the $ab$ plane, separated by nonmagnetic TlS$_6$ layers [Fig.~\ref{structure}(a,b)]. Within each layer, the Yb ions form an ideal triangular lattice with a NN Yb–Yb separation of $\sim 3.93$~\AA. The NNN distance within the plane is $\sim 6.8$~\AA, providing the relevant length scale for NNN exchange interactions ($J_2$). The YbS$_6$ octahedra are only weakly distorted. The S atoms form two equilateral triangles with edge lengths of $\sim 3.93$~\AA, along with six slightly distorted isosceles triangles with edge lengths of $\sim 3.93$~\AA\ and $\sim 3.71$~\AA. Along the $c$ axis, three such YbS$_6$ layers are stacked within the unit cell in an ABCA-type sequence, with an interlayer Yb–Yb separation of $\sim 7.82$~\AA. The large ratio between the interlayer and in-plane Yb--Yb separations highlights the quasi-two-dimensional nature of the magnetic lattice. Consequently, the interlayer magnetic coupling is expected to be significantly weaker than the dominant in-plane interactions.

\begin{figure*}[!htb]
  \includegraphics[scale=0.7]{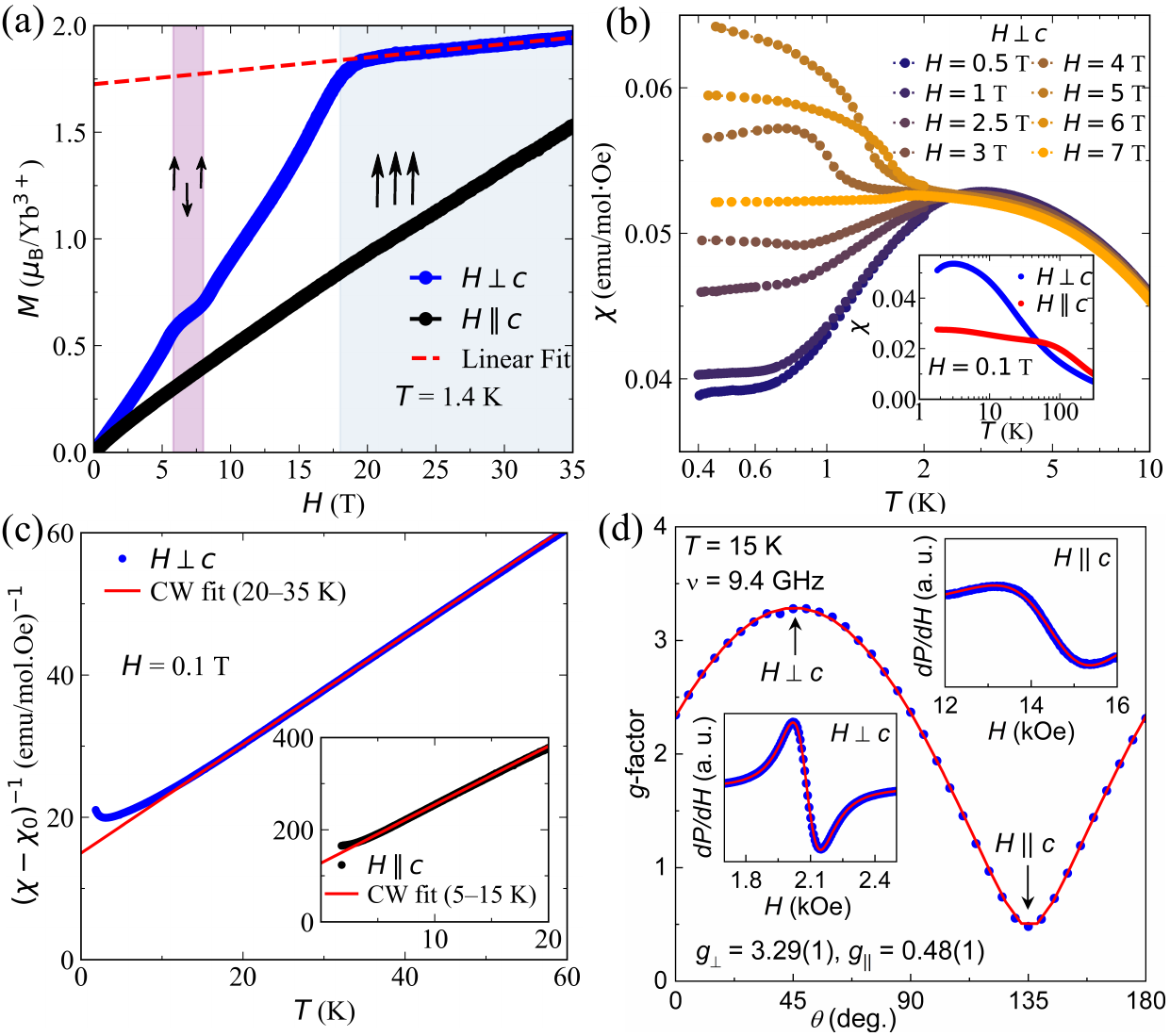}
\caption{DC magnetization and ESR of TlYbS$_2$.
(a) Isothermal magnetization $M(H)$ at $T = 1.4$~K for $H \perp c$ (blue) and $H \parallel c$ (black). For $H \perp c$, a clear saturation is observed near $H_{\rm sat} \simeq 18$~T, with a high-field linear fit (red dashed line) yielding $M_{\rm sat} \simeq 1.7\,\mu_{\rm B}$ per Yb$^{3+}$. A 1/3 plateau near $M_{\rm sat}/3$, characteristic of the up–up–down configuration in TLAFs, is evident. For $H \parallel c$, no saturation is observed up to $35$~T. (b) Temperature dependence of the DC magnetic susceptibility $\chi(T)$ for $H\perp c$ under various applied fields. A field-induced anomaly emerges near $2.5$~T. Inset: $\chi(T)$ measured at $H=0.1$~T for $H\perp c$ and $H\parallel c$.
(c) Low-temperature Curie--Weiss analysis at $H=0.1$~T. The main panel shows the fit for $H\perp c$ over $20$--$35$~K, while the inset shows the corresponding fit for $H\parallel c$ over $5$--$15$~K.
(d) Angular dependence of the Yb$^{3+}$-ESR at $T=15$~K and $\nu=9.4$~GHz, yielding strongly anisotropic $g$ factors. The insets show ESR signals at $15$~K, corresponding to Lorentz lines for both orientations.}
\label{DC}
\end{figure*}

\subsection{DC Magnetization and ESR}
DC magnetization measurements were carried out to characterize the magnetic anisotropy and estimate the magnetic energy scale of TlYbS$_2$. Fig.~\ref{DC}(a) shows the isothermal magnetization, $M(H)$, measured at $T=1.4$~K for magnetic fields applied perpendicular ($H\perp c$) and parallel ($H\parallel c$) to the crystallographic $c$ axis. For $H\perp c$, the magnetization increases nonlinearly with field and approaches saturation near $H_{\rm sat}^{\perp}\simeq18$~T. In contrast, for $H\parallel c$, $M(H)$ remains nearly linear up to 35~T, with no indication of saturation. This pronounced difference directly establishes an easy-plane magnetic anisotropy in TlYbS$_2$. A clear $1/3$ magnetization plateau is observed between $5$ and $8$~T for $H \perp c$, consistent with the up–up–down spin configuration expected for TLAFs~\cite{balents2010spin,chen2013ground,honecker2004magnetization,syromyatnikov2023unusual}. Above $H_{\rm sat}^{\perp}$, the magnetization continues to increase linearly, indicating a sizable Van Vleck contribution. A linear fit to the high-field region yields $\chi_{\rm VV}^{\perp} \simeq 3.5 \times 10^{-3}$ emu/mol·Oe. After subtracting this contribution, the saturation magnetization is $M_{\rm sat}^{\perp} \simeq 1.7~\mu_{\rm B}$, consistent with a $J_{\rm eff}=1/2$ ground state and $g_{\perp} \simeq 3.4$. The absence of saturation for $H \parallel c$ prevents a reliable estimate of $\chi_{\rm VV}^{\parallel}$. The in-plane saturation field of TlYbS$_2$ is comparatively high among the studied Yb-based delafossites~\cite{li2024thermodynamics,lee2024magnetic,ranjith2019anisotropic,xing2021synthesis,xie2023complete}, although it remains below the $\sim22$~T saturation field reported for TlYbSe$_2$~\cite{belbase2026finite}.

The strong anisotropy observed in $M(H)$ is also evident in the low-field magnetic susceptibility, $\chi(T)$, shown in the inset of Fig.~\ref{DC}(b). The temperature dependence of $\chi(T)$ was measured at $H=0.1$~T for both $H\perp c$ and $H\parallel c$. No signature of long-range magnetic order is observed down to approximately $2$~K. For $H\perp c$, $\chi(T)$ exhibits a broad maximum near $\sim3$~K, consistent with the development of short-range magnetic correlations expected in frustrated TLAFs. A pronounced difference between $\chi_{\perp}$ and $\chi_{\parallel}$ persists over a broad temperature range, further demonstrating the strongly anisotropic magnetic response of TlYbS$_2$.

To further investigate the field-induced behavior observed in $M(H)$ for $H\perp c$, we measured $\chi(T)$ down to 0.4~K, as shown in the main panel of Fig.~\ref{DC}(b). At low fields, no clear signature of a magnetic phase transition is observed down to the base temperature. At an applied field of approximately $2.5$~T, a weak anomaly emerges in $\chi(T)$ and becomes more pronounced with further increasing field. The corresponding transition temperature shifts to higher values with increasing field up to $\sim 7$~T, indicating the stabilization of a field-induced phase. This evolution closely follows the field scales identified in the $M(H)$ measurements [Fig.~\ref{DC}(a)].

To obtain approximate estimates of the low-temperature effective moments and magnetic interaction scales, we next analyze the inverse susceptibility. At high temperatures, $1/\chi(T)$ follows Curie–Weiss (CW) behavior, yielding an effective moment consistent with the theoretical value for free Yb$^{3+}$ ions with $J=7/2$ and $g=8/7$ (see SM ~\cite{supp}). However, due to strong spin–orbit coupling and crystal electric field effects, the low-temperature physics is governed by a Kramers doublet, allowing an effective $J_{\rm eff}=1/2$ description~\cite{bordelon2019field,ranjith2019anisotropic,hester2019novel,paddison2017continuous,li2015gapless,Arjun014013,Arjun224415}. Therefore, we perform modified CW fits in the low-temperature regime using $\chi(T)=\chi_0 +C/(T-\theta_{\rm CW})$, where $\chi_0$ is a temperature-independent Van Vleck contribution, $\theta_{\rm CW}$ is the Curie-Weiss temperature, and $C$ is the Curie constant. For $H \perp c$, $\chi_0^{\perp}$ is fixed to the value obtained from $M(H)$, while $\chi_0^{\parallel}$ is treated as a free parameter. The fits are performed over $20$–$35$ K for $H \perp c$ and $5$–$15$ K for $H \parallel c$ [Fig.~\ref{DC}(c)]. The extracted parameters are $\mu_{\rm eff}^{\perp} \simeq 3.2~\mu_{\rm B}$ and $\mu_{\rm eff}^{\parallel} \simeq 0.9~\mu_{\rm B}$, with $\theta_{\rm CW}^{\perp} \simeq -19.5$~K and $\theta_{\rm CW}^{\parallel} \simeq -10$~K. The temperature-independent susceptibility obtained from the fits is $\chi_0^{\perp} \simeq 3.5 \times 10^{-3}$~emu/mol$\cdot$Oe and $\chi_0^{\parallel} \simeq 21 \times 10^{-3}$~emu/mol$\cdot$Oe. These relatively large values indicate a significant Van Vleck contribution for both field directions. In particular, $\chi_0^{\parallel}$ is significantly larger than $\chi_0^{\perp}$, over a broad temperature range [Fig.~\ref{DC}(b) inset]. 
Using $\mu_{\rm eff}=g\sqrt{J(J+1)}\mu_{\rm B}$ with $J=1/2$, we obtain $g_{\perp}\simeq3.7$ and $g_{\parallel}\simeq1.04$. The negative $\theta_{\rm CW}$ values confirm dominant antiferromagnetic interactions. Notably, the Curie–Weiss temperatures for TlYbS$_2$ are larger in magnitude than those reported for other Yb-based delafossites~\cite{baenitz2018naybs,ranjith2019anisotropic,xing2021synthesis,bordelon2019field}, indicating an enhanced exchange energy scale in this system. Using the relation, $\theta_{\rm CW}^{\rm \perp}=(-3/2)J_{\perp}$ and  $\theta_{\rm CW}^{\rm \parallel}=(-3/2)J_{z}$ for $H \perp c$ and $H \parallel c$, respectively~\cite{schmidt2017frustrated}, we obtain rough estimates of $J_{\perp} \simeq 13$~K and $J_{z} \simeq 6.67$~K. 

To directly probe the magnetic anisotropy, we performed ESR measurements at 15~K. The angular dependence of the ESR $g$ factor is shown in Fig.~\ref{DC}(d), with representative ESR spectra for $H \perp c$ and $H \parallel c$ displayed in the inset. Both spectra are well described by Lorentzian line shapes. The angular dependence of the $g$ factor is well described by $g(\theta)=\sqrt{g_{\parallel}^{2}\cos^{2}\theta+g_{\perp}^{2}\sin^{2}\theta}$, yielding $g_{\perp}=3.29(1)$ and $g_{\parallel}=0.48(1)$. The resulting anisotropy ratio, $g_{\perp}/g_{\parallel}\approx 6.9$, demonstrates a highly anisotropic $g$ tensor and provides an independent microscopic measure of the pronounced easy-plane magnetic anisotropy inferred from the bulk magnetization measurements. This anisotropy is significantly larger than that reported for NaYbSe$_2$ ($g_{\perp}/g_{\parallel}\approx 3.1$)~\cite{ranjith2019anisotropic} and NaYbS$_2$ ($g_{\perp}/g_{\parallel}\approx 5.6$)~\cite{baenitz2018naybs}, but remains smaller than those observed in TlYbSe$_2$ ($g_{\perp}/g_{\parallel}\approx 9.2$)~\cite{fujii2025tlybse} and CsYbSe$_2$ ($g_{\perp}/g_{\parallel}\approx 10.8$)~\cite{xie2023complete}. The ESR-derived value $g_{\perp}=3.29(1)$ is in excellent agreement with $g_{\perp}\simeq3.4$ independently estimated from the saturation magnetization [Fig.~\ref{DC}(a)], and is also reasonably consistent with $g_{\perp}\simeq3.7$ obtained from the low-temperature CW analysis [Fig.~\ref{DC}(c)]. This consistency provides further support for the effective $J_{\rm eff}=1/2$ description of the ground-state Kramers doublet in TlYbS$_2$. In contrast, the susceptibility-derived value $g_{\parallel}\simeq1.04$ is noticeably larger than the ESR value $g_{\parallel}=0.48(1)$. This discrepancy likely reflects the substantial Van Vleck contribution for $H\parallel c$, which complicates the extraction of the intrinsic Curie susceptibility. The large $g_{\perp}$ is consistent with the tendency toward magnetization saturation for $H\perp c$, whereas the much smaller $g_{\parallel}$ is consistent with the absence of saturation for $H\parallel c$ within the experimentally accessible field range. 

In addition, the ESR linewidth is substantially broader for $H \parallel c$ than for $H \perp c$. Although part of this broadening can be attributed to the pronounced $g$ tensor anisotropy, the linewidth anisotropy appears significantly stronger than expected from the resonance-field anisotropy alone, which may reflect additional anisotropic broadening mechanisms beyond the $g$ factor anisotropy. Additional temperature-dependent ESR measurements are presented in the SM~\cite{supp}. The ESR intensity exhibits CW-like behavior with negative Weiss temperatures, consistent with dominant antiferromagnetic correlations. Furthermore, an Orbach analysis of the ESR linewidth yields activation energies of approximately $270$~K for $H\perp c$, indicating a well-isolated ground-state Kramers doublet. Due to the extremely broad signal, which on increasing temperature partially exceeds the accessible field range, a reliable value for the activation energy cannot be determined for $H\parallel c$.

\begin{figure*}[!htb]
  \includegraphics[width=\linewidth]{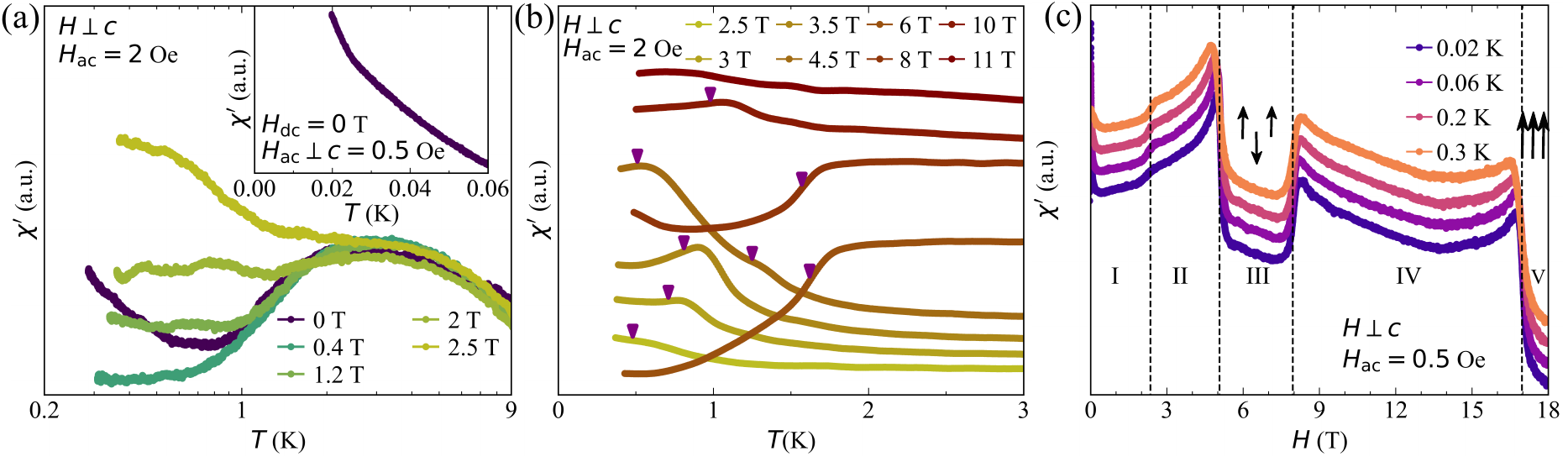}
\caption{AC susceptibility of TlYbS$_2$ for $H \perp c$.
(a) Temperature dependence of the real part of the AC susceptibility, $\chi'(T)$, measured from $0.3$ to $9$~K under magnetic fields up to $2.5$~T, showing the onset of a field-induced magnetic transition near $2.5$~T. Inset: $\chi'(T)$ measured at zero DC field down to $20$~mK. The susceptibility increases continuously upon cooling without developing a discernible peak. (b) Evolution of $\chi'(T)$ under higher magnetic fields, illustrating the emergence and systematic evolution of magnetic transitions with increasing field. Purple arrows mark the transition temperatures. Curves are vertically offset (non-uniformly) for clarity; the original, unshifted data are shown in the Supplementary Material~\cite{supp}. (c) Field-dependent $\chi'(H)$ data measured at various temperatures from $20$~mK to $300$~mK, highlighting multiple field-induced transitions, with phase boundaries denoted by dotted black lines. The background signal has not been subtracted, and curves are vertically offset (uniformly) for clarity.}

\label{AC}
\end{figure*}

\subsection{AC Susceptibility}

The low-temperature magnetic response for $H\perp c$, beyond the range accessible in DC measurements, was investigated using AC susceptibility measurements down to $\sim20$~mK. 
No spin freezing anomaly was observed in the single crystals down to 20 mK. The real part of the AC susceptibility, $\chi'(T)$, increases continuously upon cooling down to $20$~mK without developing a discernible peak, as shown in the inset of Fig.~\ref{AC}(a). This behavior contrasts with the frequency-dependent peaks observed in TlYbSe$_2$ and NaYbSe$_2$ in the $\sim30- 50$~mK temperature range~\cite{belbase2026finite,scheie2024spectrum}. With increasing magnetic field, $\chi'(T)$ develops a clear anomaly near $2.5$~T [Fig.~\ref{AC}(a) and Fig.~\ref{AC}(b)], indicating the onset of a field-induced phase consistent with the DC magnetization measurements. This anomaly evolves into a well-defined peak, which can be tracked up to $\sim 10$~T and marks the corresponding phase boundary. At higher fields ($\gtrsim 11$~T), the overall AC signal decreases, and the associated anomalies become progressively weaker, approaching the experimental resolution. 
The transition temperature reaches a maximum of $\sim 1.7$~K near $7$~T, which is comparable to values reported for TlYbSe$_2$~\cite{belbase2026finite,fujii2025tlybse}, but higher than those observed in other delafossite systems of this family~\cite{lee2024magnetic,ranjith2019anisotropic}.

The isothermal susceptibility, $\chi'(H)$, shown in Fig.~\ref{AC}(c), provides complementary insight into the field-driven evolution of magnetic phases. A clear anomaly appears near $ 2.3$~T, marking the onset of a field-induced phase. With further increase in field, $\chi'(H)$ rises and reaches a maximum around $5$~T, indicating a transition into a distinct magnetic state. Upon increasing the field further, $\chi'(H)$ decreases and develops a minimum in the range $\sim 5.5$–$7.5$~T. This field window coincides with the region where the transition temperature attains its maximum in the $\chi'(T)$ data [Fig.~\ref{AC}(b)] and corresponds to the $M_{\rm sat}/3$ plateau observed in $M(H)$ [Fig.~\ref{DC}(a)]. For $H \gtrsim 7.5$~T, $\chi'(H)$ begins to increase again, consistent with the emergence of a higher-field phase whose transition boundaries shift to lower temperatures. At higher fields, the spins progressively align along the applied field direction, becoming fully polarized near $17$~T, as reflected in the pronounced suppression of $\chi'(H)$.

\begin{figure*}[!htb]
  \includegraphics[scale=0.7]{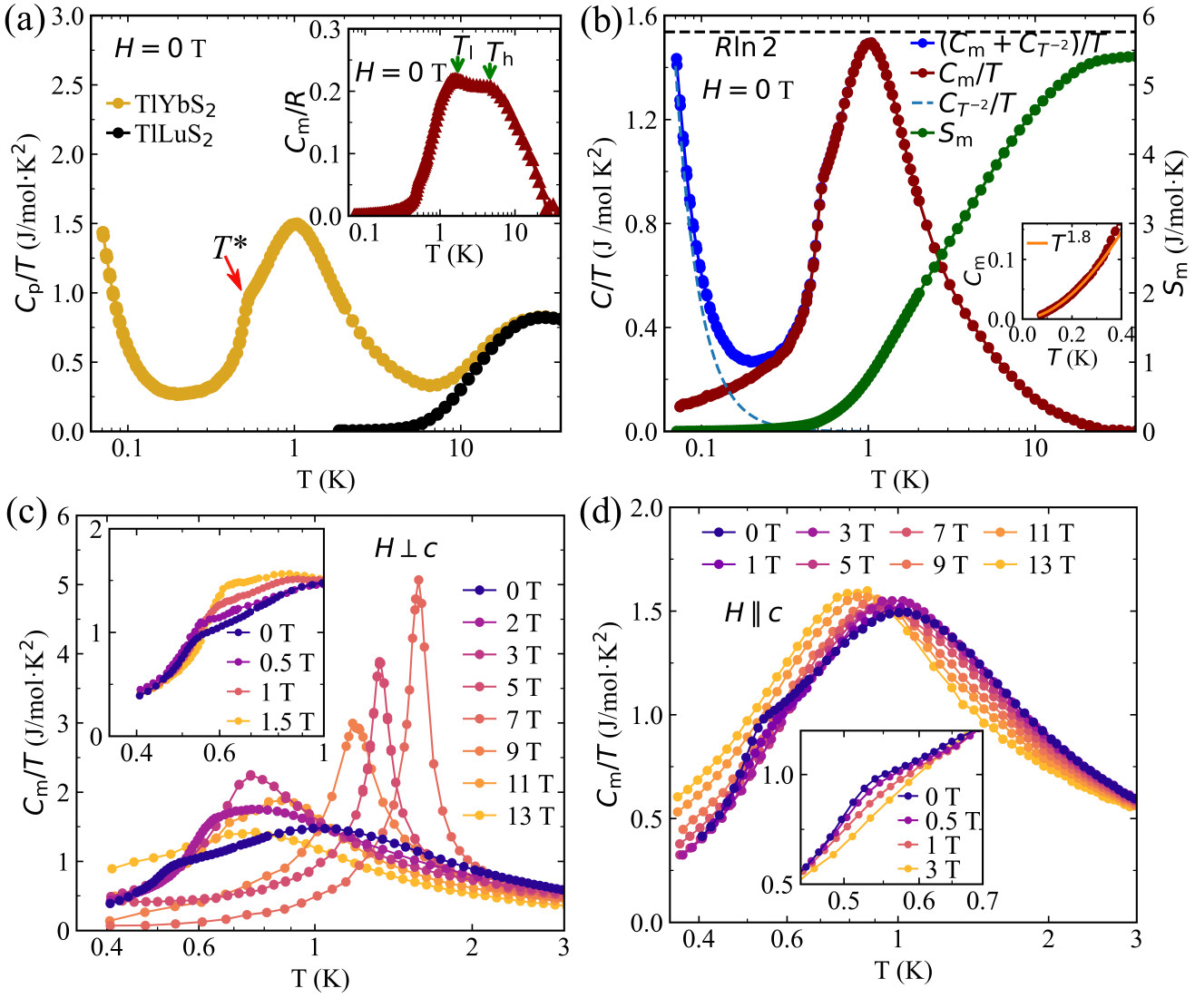}
\caption{
Specific heat of TlYbS$_2$.
(a) Temperature dependence of $C_{\rm p}/T$ at zero field for TlYbS$_2$ and its isostructural nonmagnetic analog TlLuS$_2$. The anomaly at $\sim530$~mK is marked as $T^*$. Inset: magnetic contribution to the specific heat, $C_{\rm m}/R$, obtained after subtracting the phonon contribution and the fitted inverse-square contribution, showing two broad maxima centered near $T_{\rm l}\simeq1.7$~K and $T_{\rm h}\simeq4.6$~K.
(b) Temperature dependence of the phonon-subtracted specific heat, $(C_{\rm m}+C_{T^{-2}})/T$ (blue), the fitted inverse-square contribution, $C_{T^{-2}}/T$ (dashed blue), and the resulting magnetic contribution, $C_{\rm m}/T$ (dark red), together with the magnetic entropy $S_{\rm m}(T)$ (green, right $y$-axis). The horizontal dashed line corresponds to the entropy of a Kramers doublet, $R\ln2$. Inset: low-temperature $C_{\rm m}$ after subtraction of the fitted inverse-square contribution. The solid orange line represents a power-law fit, $C_{\rm m}\propto T^{1.8}$.
(c) $C_{\rm m}/T$ measured under magnetic fields up to $13$~T applied perpendicular to the $c$ axis ($H\perp c$). A field-induced magnetic transition emerges above $\sim2.5$~T, becomes most pronounced near $\sim7$~T, and is progressively suppressed at higher fields. Inset: low-field $C_{\rm m}/T$ highlighting the evolution of the weak zero-field anomaly under $H\perp c$.
(d) $C_{\rm m}/T$ measured under magnetic fields up to $13$~T applied parallel to the $c$ axis ($H\parallel c$). Inset: low-field $C_{\rm m}/T$ showing the evolution of the zero-field anomaly under $H\parallel c$.}
\label{HC}
\end{figure*}

\subsection{Specific Heat}
To further probe the low-temperature magnetic behavior of TlYbS$_2$, we performed specific-heat measurements down to $70$~mK. Figure~\ref{HC}(a) shows the zero-field $C_{\rm p}/T$ of TlYbS$_2$ together with that of the isostructural nonmagnetic analog TlLuS$_2$, which was used to estimate the phonon contribution, $C_{\rm ph}(T)$~\cite{supp}. 
The magnetic contribution, $C_{\rm m}(T)$, shown in the inset of Fig.~\ref{HC}(a), was obtained after subtracting both the phonon and empirical inverse-square contributions, with the latter determined from the analysis described below.


The resulting $C_{\rm m}(T)$ exhibits broad maxima extending over $\sim 1-10$~K. Such behavior is commonly observed in low-dimensional frustrated magnets and is indicative of a crossover into a regime governed by strong quantum fluctuations~\cite{yamashita2008thermodynamic,li2015gapless}. A similar broad maximum has been reported in several members of the delafossite family, where it is often attributed to the superposition of two broad features associated with distinct exchange energy scales~\cite{ranjith2019anisotropic,bordelon2019field}, consistent with theoretical predictions for fully frustrated two-dimensional triangular and kagome antiferromagnets~\cite{ishida1997low,elser1989nuclear}. For spin-$1/2$ TLAFs, the characteristic crossover temperatures are expected at
$T_l/J_1 \simeq 0.2$ and $T_h/J_1 \simeq 0.55$
\cite{chen2019two}. The observed broad maxima occur at reduced
temperatures comparable to those expected for frustrated TLAFs, providing support for interpreting them as crossover
features associated with the development of short-range magnetic correlations. The observed peak magnitude, $C_{\rm m}^{\rm max} \simeq 0.22R$,
is also consistent with theoretical expectations for
spin-$1/2$ TLAFs
\cite{elstner1993finite,bernu2001specific}.

In addition to the broad maximum, a weak anomaly is observed near $\sim530$~mK in $C_{\rm p}/T$ [Fig.~\ref{HC}(a)].
This feature is reproducible in measurements on crystals grown from different batches (see SM~\cite{supp}), supporting its intrinsic origin. We tentatively associate this anomaly with the onset of a magnetically ordered state. 
At the lowest temperatures, the phonon-subtracted specific heat contains an additional upturn that is well described by an inverse-square contribution. To determine the temperature dependence of the intrinsic magnetic specific heat, we fit the low-temperature data between $0.07$ and $0.30$~K using
$C_{\rm p}-C_{\rm ph}=\gamma T^{b}+\alpha/T^{2}$.
Here, $C_{T^{-2}}=\alpha/T^{2}$ represents an empirical inverse-square contribution to the low-temperature specific heat, while $C_{\rm m}=\gamma T^{b}$ describes the intrinsic magnetic contribution. The fit yields $\alpha=4.820(20)\times10^{-4}$~J\,K/mol and $b=1.784(17)$, with $\gamma=0.744(18)$~J\,mol$^{-1}$\,K$^{-(b+1)}$. The corresponding low-temperature fit is shown in the SM~\cite{supp}. The extracted parameters show only weak dependence on the fitting range (see SM~\cite{supp}), indicating that the observed nearly quadratic low-temperature behavior is robust. Furthermore, the obtained $\alpha$ value is comparable to those reported for TlYbSe$_2$ and other Yb-based delafossite compounds~\cite{belbase2026finite}, supporting the validity of the adopted low-temperature decomposition. The magnetic specific heat follows approximately $C_{\rm m}\propto T^{1.8}$ between 0.07 and 0.30~K, close to a quadratic temperature dependence.
The magnetic entropy, $S_{\rm m}$, obtained by integrating $C_{\rm m}/T$ from $70$~mK, reaches approximately $94\%$ of $R\ln2$ by $30$~K [Fig.~\ref{HC}(b)], close to the value expected for a well-isolated Kramers doublet ground state and an effective $J_{\rm eff}=1/2$ description of Yb$^{3+}$. The unrecovered entropy may originate from contributions below $70$~mK and/or uncertainties associated with the determination of $C_{\rm m}$. Notably, less than $1\%$ of the total magnetic entropy is released in the vicinity of the weak anomaly near $530$~mK, indicating that the transition emerges from a strongly fluctuating magnetic background.

To investigate the evolution of the low-temperature anomaly under an applied magnetic field, we performed temperature-dependent specific heat measurements at different applied magnetic fields with the field oriented both perpendicular ($H{\perp}c$) and parallel ($H{\parallel}c$) to the $c$ axis, as shown in Fig.~\ref{HC}(c,d). For $H{\perp}c$, the anomaly remains robust up to approximately 2~T and may even exhibit a slight enhancement on the higher-temperature side [inset of Fig.~\ref{HC}(c)]. For $H{\perp}c\gtrsim2.5$~T, a field-induced phase emerges, seen as a pronounced peak in $C_{\rm m}/T$ [Fig.~\ref{HC}(c)], consistent with the AC and DC magnetization results, which also show signatures of field-induced ordering. With increasing field, this transition becomes sharper and shifts to higher temperature, reaching a maximum near $7$~T for $H\perp c$, before weakening and becoming unresolved near $13$~T.

For $H\parallel c$, by contrast, the anomaly shows only a minimal field-induced shift and is suppressed near $3$~T, without the emergence of a comparable field-induced ordered phase [inset of Fig.~\ref{HC}(d)], consistent with the absence of field-induced transitions in the $M(H)$ data [Fig.~\ref{DC}(a)]. This suppression field coincides with the critical field $H_c \approx 3$~T at which independent thermal-conductivity and NMR measurements report a confinement--deconfinement transition from the ordered state into a gapless quantum spin liquid with a spinon Fermi surface for $H\parallel c$~\cite{hosoi2026continuous}.

Overall, the contrasting field response for the two orientations highlights the strong magnetic anisotropy of TlYbS$_2$: for $H{\perp}c$, the anomaly evolves into a field-induced ordered phase, whereas for $H\parallel c$ it instead gives way to a fluctuation-dominated, quantum-disordered regime. This anisotropy is consistent with the ESR results [Fig.~\ref{DC}(d)] and $M(H)$ data [Fig.~\ref{DC}(a)] and underscores the crucial role of spin anisotropy and field orientation in determining the magnetic ground state of TlYbS$_2$.

\begin{figure}[!htb]
  \includegraphics[scale=0.7]{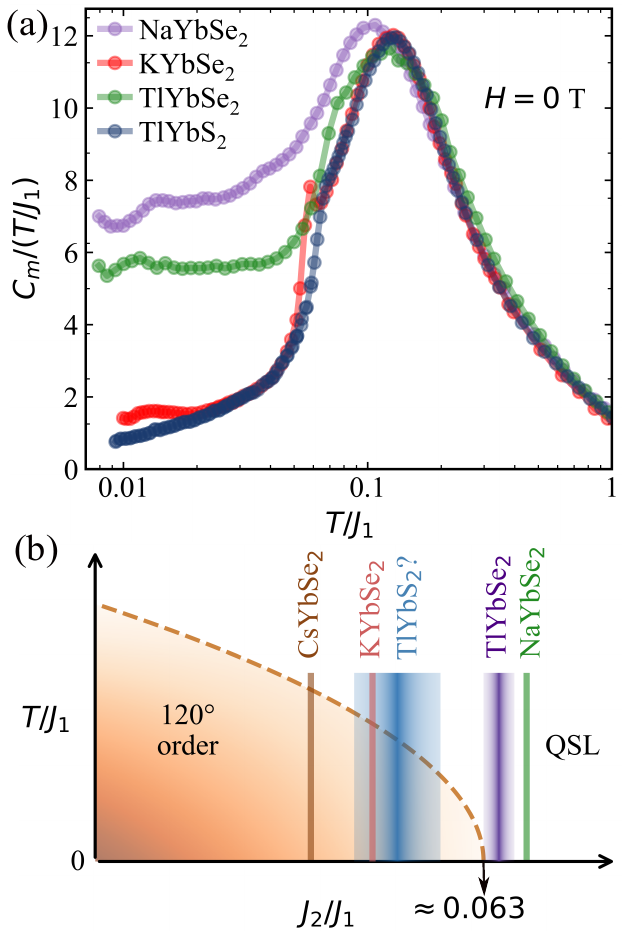}
\caption{Scaled magnetic specific heat and schematic $J_1$--$J_2$ phase diagram for Yb-based triangular-lattice delafossites. (a) Comparison of the normalized magnetic specific heat,
$C_{\rm m}/(T/J_1)$, as a function of reduced temperature
$T/J_1$ for TlYbS$_2$ and representative members of the
Yb-based triangular-lattice delafossite family. The data for
NaYbSe$_2$, TlYbSe$_2$, and KYbSe$_2$ are taken from
Refs.~\cite{ranjith2019anisotropic,belbase2026finite,scheie2024spectrum}.
As a reliable microscopic determination of $J_1$ and $J_2$ is presently unavailable for TlYbS$_2$, the scaling, performed using an approximate value of $J_1 = 8$~K, should be viewed as qualitative and intended only to facilitate comparison with other Yb-based delafossites. (b) Schematic phase diagram of the spin-$1/2$ triangular-lattice
$J_1$--$J_2$ Heisenberg model in the isotropic exchange limit.
The diagram is shown only as a qualitative guide to the
evolution from the $120^\circ$-ordered state to the quantum
spin-liquid (QSL) regime, with the proposed critical point
located near $J_2/J_1 \approx 0.063$. The phase boundary and
the placement of previously studied compounds are adapted
from Refs.~\cite{scheie2024nonlinear,scheie2024spectrum,belbase2026finite}, while the positions of TlYbS$_2$ and TlYbSe$_2$ are schematic.} 
\label{HC_comparison}
\end{figure}

\section{Discussion} 

A broad, kink-like anomaly is observed in the specific heat near $530$~mK; however, no corresponding feature is resolved in either the DC magnetization or AC susceptibility. Similar behavior has also been reported in KYbSe$_2$, suggesting a common phenomenology among Yb-based triangular-lattice delafossites~\cite{scheie2024spectrum}. The anomaly in TlYbS$_2$ is smeared compared with the relatively sharp feature observed in KYbSe$_2$. Nevertheless, it is consistently reproduced in measurements
on crystals grown from different batches (see SM~\cite{supp}), supporting its intrinsic origin. The absence of a corresponding feature in the magnetization and AC susceptibility data means that the nature of this anomaly cannot be
established from the present bulk measurements alone. This picture is consistent with independent measurements on TlYbS$_2$, which report an ordered state below $\approx530$~mK emerging from a strongly fluctuating background, with low-energy fluctuations persisting within the ordered phase~\cite{hosoi2026continuous}. Interestingly, below $300$~mK, the magnetic specific heat follows approximately $C_{\rm m}\propto T^{1.8}$, which differs from the nearly linear-in-$T$ magnetic specific heat observed in TlYbSe$_2$\cite{belbase2026finite}.

To place TlYbS$_2$ in the broader context of the Yb-based delafossite family, Fig.~\ref{HC_comparison}(a) compares the scaled magnetic specific heat of several representative compounds using the nearest-neighbor exchange interaction $J_1$ as the characteristic energy scale. All four compounds exhibit a broad maximum near $T/J_1 \sim 0.1$, reflecting the development of short-range magnetic correlations. Clear differences emerge at lower temperatures. NaYbSe$_2$ and TlYbSe$2$ exhibit an approximately linear temperature dependence of $C_{\rm m}$ without a clear specific-heat anomaly associated with long-range magnetic ordering. In contrast, KYbSe$_2$ shows a sharp feature near $290$~mK. TlYbS$_2$ exhibits a similar, albeit smeared, anomaly at a comparable reduced temperature and an approximately quadratic temperature dependence at low temperatures. In the low-temperature regime, the magnitude of the scaled magnetic specific heat, $C_{\rm m}/(T/J_1)$, is largest for NaYbSe$_2$, followed by TlYbSe$_2$, whereas KYbSe$_2$ and TlYbS$_2$ exhibit comparable but lower values over their common reduced-temperature range. Based on the overall scaled specific-heat comparison and character of the anomaly, we tentatively place TlYbS$_2$ near the boundary between the $120^\circ$-ordered and QSL regimes in the schematic $J_1$--$J_2$ phase diagram shown in Fig.~\ref{HC_comparison}(b). This placement
should be regarded as qualitative, as a reliable microscopic determination of $J_1$ and $J_2$ is not yet available for TlYbS$_2$.

The resulting $H$–$T$ phase diagram, constructed from $\chi'(T)$, $\chi'(H)$, $M(H)$, $\chi(T)$, and $C_{\rm p}(T)$ data for in-plane applied field ($H \perp c$), is presented in Fig.~\ref{structure}(d). As the magnetic field increases, TlYbS$_2$ evolves from a low-field correlated regime through a sequence of field-induced magnetic states. Phase~I denotes the low-field regime containing the weak thermodynamic anomaly near $T^*\simeq530$~mK. Phase~II, which emerges above approximately $2.3$~T, is tentatively assigned to an oblique $120^\circ$ (Y-coplanar) state. Phase~III, between approximately $5.1$ and $8$~T, is assigned to an up--up--down (UUD) collinear state, consistent with the suppression of $\chi'(H)$ and the $1/3$ magnetization plateau in $M(H)$. Experimentally, such UUD phases have been observed in several TLAF compounds \cite{ranjith2019anisotropic,lee2014magnetic,lee2024magnetic,tsujii2007thermodynamics,smirnov2007triangular,lee2017magnetic,gao2022spin,hwang2012successive}. Phase~IV is tentatively assigned to a $2{:}1$ canted spin configuration~\cite{seabra2011phase}. Phase~V corresponds to the high-field polarized regime reached above $\sim 17$~T. Within the Yb-based triangular-lattice delafossite family, TlYbS$_2$ occupies a comparatively high magnetic energy scale. While NaYbO$_2$, NaYbS$_2$, NaYbSe$_2$, KYbSe$_2$, RbYbSe$_2$, and CsYbSe$_2$ reach the fully spin-polarized state at relatively lower magnetic fields~\cite{bordelon2019field,lee2024magnetic,ranjith2019anisotropic,xing2021synthesis,xie2023complete}, TlYbS$_2$ requires an in-plane field of nearly $17$~T for full polarization. Its magnetic energy scale is, however, somewhat smaller than that of TlYbSe$_2$, which reaches full polarization only near $22$~T~\cite{belbase2026finite}. Despite these differences in the absolute field scales, the overall sequence of field-induced phases is broadly similar to that observed in other Yb-based triangular-lattice delafossites.

\section{CONCLUSION}

In conclusion, we have synthesized and characterized high-quality single crystals of TlYbS$_2$, an effective spin-$1/2$ triangular-lattice antiferromagnet in the Yb-based delafossite family. Magnetization, susceptibility, and ESR measurements consistently reveal strong easy-plane magnetic anisotropy, with $g_{\perp}\approx3.3$ and $g_{\parallel}\approx0.5$, and support a well-isolated Kramers-doublet ground state. The thermodynamic anomaly near $\sim530$~mK is consistent with the onset of a weakly ordered state developing in the presence of strong magnetic fluctuations~\cite{hosoi2026continuous}. Crucially, a complementary study for $H\parallel c$, performed on crystals from the same series as those used in this work, reports an ordered state emerging from a strongly fluctuating background and a confinement--deconfinement transition into a gapless quantum spin liquid above $H_c\approx3$~T~\cite{hosoi2026continuous}, whereas for $H\perp c$ we observe a distinct sequence of field-induced magnetic phases beginning near $\sim2.5$~T. These results suggest that TlYbS$_2$ lies close to the boundary between weak magnetic order and a quantum-disordered regime. Future neutron-scattering measurements and a reliable microscopic determination of $J_1$ and $J_2$ will be important for establishing the nature of the weakly ordered state, testing its location relative to the proposed ordered-to-QSL boundary, and clarifying the low-energy excitation spectrum.

\acknowledgments 
\noindent We thank S. Hosoi, Y. Matsuda, A. Tsirlin, and J. Xing for the fruitful discussions and valuable suggestions. The research as a whole is supported by the U.S. Department of Energy – Office of Science, Basic Energy Sciences (Grant No. DE-SC0022986), under the project \textit{``Seeking quasiparticles in perturbed matter from low-energy spin dynamics"}. A.U. acknowledges financial support from the Department of Science and Technology (DST), Government of India, through the DST Inspire Faculty Fellowship (Ref. No. DST/INSPIRE/04/2019/001664). The work in Augsburg was funded by the Deutsche Forschungsgemeinschaft (DFG, German Research Foundation) through Project No. TRR 360--492547816 (Subproject B3). We gratefully acknowledge the Spin-Lab facility at the Birck Nanotechnology Center, Purdue University, for providing access to its measurement facilities. Part of this work was conducted at the National High Magnetic Field Laboratory, supported by the National Science Foundation (Cooperative Agreement No. DMR-2128556) and the State of Florida.\\

\bibliographystyle{apsrev4-2}
\bibliography{references_v1}

\end{document}